\documentclass[12pt,a4paper]{iopart}

\usepackage{bbm}
\usepackage{graphicx}
\usepackage{oldgerm}
\usepackage{mathpak}
\usepackage{amssymb}

\newcommand{\bra}[1]{\ensuremath{\left\langle #1\right|}}
\newcommand{\ket}[1]{\ensuremath{\left|#1\right\rangle}}

\begin{document}

\title{A composite parameterization of unitary groups, density matrices and subspaces}

\author{Christoph Spengler$^1$, Marcus Huber$^1$ and Beatrix C. Hiesmayr$^{1,2}$}
\address{$^1$Faculty of Physics, University of Vienna, Boltzmanngasse 5, 1090 Vienna, Austria}
\address{$^2$Faculty of Physics, Sofia University, 5 James Bourchier Blvd., 1164 Sofia, Bulgaria}
\ead{\mailto{Christoph.Spengler@univie.ac.at}, \mailto{Marcus.Huber@univie.ac.at}, \mailto{Beatrix.Hiesmayr@univie.ac.at}}

\begin{abstract}
Unitary transformations and density matrices are central objects in quantum physics and various tasks require to introduce them in a parameterized form. In the present article we present a parameterization of the unitary group $\mathcal{U}(d)$ of arbitrary dimension $d$ which is constructed in a composite way. We show explicitly how any element of $\mathcal{U}(d)$ can be composed of matrix exponential functions of generalized anti-symmetric $\sigma$-matrices and
one-dimensional projectors. The specific form makes it considerably easy to identify and discard redundant parameters in several cases. In this way, redundancy-free density matrices of arbitrary rank $k$ can be formulated. Our construction can also be used to derive an orthonormal basis of any $k$-dimensional subspaces of $\mathbb{C}^d$ with the minimal number of parameters. As an example it is shown that this feature leads to a significant reduction of parameters in the case of investigating distillability of quantum states via lower bounds of an entanglement measure (the $m$-concurrence).
\end{abstract}


\maketitle

\section{Introduction}
In quantum information many quantities or properties of quantum systems are related to optimization problems that necessitate to vary over the set of all unitary transformations, density matrices or subspaces. Such problems arise, for instance, in the detection and quantification of entanglement (see \cite{GuehneDet,HorodeckiEnt,HHGM} and references therein), the properties of quantum states with respect to Bell inequalities \cite{Son,SHH1} or the question of distillability \cite{Horodeckibound,NPTBound}. In general, such a variation can only be done efficiently by parameterizing the object of interest. In the present work the focus is on the parameterization of the unitary group $\mathcal{U}(d)$. Density matrices and orthonormal subspaces can then easily be formulated in terms of this representation. It should be emphasized that parameterizations are in general equivalent to each other and one should not be misled to conclude that one parameterization is better than the other. However, depending on the problem a certain parameterization can be advantageous by providing more insightful results or by leading to a reduction of the number of involved parameters. The simplest parameterization of $\mathcal{U}(d)$ is the canonical parameterization which is given by $U=\mbox{exp}(i H)$, wherein the hermitian matrix $H$ can be composed of a real-valued linear combination of the $d^2-1$ generalized Gell-Mann matrices and the unity $\mathbbm{1}_d$. Despite its simple form, this parameterization has the disadvantage that there is no way how redundant parameters, which appear for some tasks can be removed beforehand (except for one parameter for the special unitary group $\mathcal{SU}(d)$). Thus the number of parameters is always $d^2$ or $d^2-1$, respectively, independent from the given problem. A different parameterization of $\mathcal{U}(d)$ and $\mathcal{SU}(d)$ in terms of generalized Euler angle was introduced by Tilma and Sudershan \cite{ToddSUN,ToddUN}. It has the advantage that it allows to eliminate redundant 'unphysical' global phases in several cases. A further parameterization of $\mathcal{U}(d)$ was recently found by Jarlskog \cite{Jarlskog1,Fujii}. Here, the unitary matrices are formulated in a recursive way, meaning that the elements of $\mathcal{U}(d)$ are expressed in terms of the elements of $\mathcal{U}(d-1)$ and a further unitary matrix containing the parameters which enables the extension to $d$. This parameterization also enables one to remove invariant phases, however, this is a nontrivial task as shown in \cite{Jarlskog2}.

In the present work we introduce a parameterization (see Sec.$2$) which is ideal for the formulation of density matrices (see Sec.$3$) and orthonormal subspaces (see Sec.$4$) since all redundancies can be easily identified and removed beforehand without having to consider fixed dimensions or special cases explicitly. Due to its concise notation and formulation in terms of matrix exponential functions of dyadic vector products, the parameterization can be easily implemented in computational programs. Moreover, the parameters have simple ranges; when gathered in a $d \times d$ matrix in a particular order, they permit an insightful interpretation of numerical computations. Concrete numerical examples, in which our parameterization reduces the number of involved variables are given in Sec.$5$. Here, we state which parameters can be discarded with respect to optimizing lower bounds of an entanglement measure (the $m$-concurrence), as well as for clarifying distillability of quantum states.
\section{Composite parameterization of the unitary group $\mathcal{U}(d)$}
Consider unitary operators $U$ acting on a Hilbert space $\mathcal{H}=\mathbb{C}^d$ with $d\geq2$ spanned by the orthonormal basis $\{ \ket{1},\ldots,\ket{d} \}$. For any $U \in \mathcal{U}(d)$ there exist $d^2$ real values $\lambda_{m,n}$ with $m,n \in \left\{ 1 , \ldots , d \right\}$ and $\lambda_{m,n} \in \left[0, 2 \pi \right]$ for $m \geq n$ and $\lambda_{m,n} \in \left[0, \frac{\pi}{2} \right]$ for $m < n$ such that $U$ equals \footnote{The sequence of the product is $\prod_{i=1}^{N}A_i=A_1 \cdot A_{2} \cdots A_N$}
\begin{align}
\label{Uc}
U_C=\left[\prod_{m=1}^{d-1} \left(\prod_{n=m+1}^{d} \mbox{exp} \left( i P_n \lambda_{n,m} \right) \mbox{exp} \left( i \sigma_{m,n} \lambda_{m,n} \right)  \right) \right] \cdot \left[ \prod_{l=1}^{d} \mbox{exp}(i P_l \lambda_{l,l})\right] \ .
\end{align}
Here, the $P_l $ are one-dimensional projectors
\begin{align}
P_l=\ket{l}\bra{l}
\end{align}
and $\sigma_{m,n}$ are the generalized anti-symmetric $\sigma$-matrices
\begin{align}
\sigma_{m,n}=-i\ket{m} \bra{n} + i \ket{n} \bra{m}
\end{align}
with $1\leq m < n \leq d$.\\
Before we prove that any unitary operator of $\mathcal{U}(d)$ can be written in the form (\ref{Uc}) let us comment on the concept behind it: Unitary transformations have the characteristic trait that they map orthonormal bases onto orthonormal bases, i.e. for a given set of vectors $ \{ \ket{1},\ldots,\ket{d} \}$ forming a orthonormal basis of a Hilbert space $\mathbb{C}^d$ the unitarily transformed set $\{ \ket{1'},\ldots,\ket{d'} \} = \{ U\ket{1},\ldots,U\ket{d}\}$ is also orthonormal. We are interested in the most general unitary operations transforming $\{ \ket{1},\ldots,\ket{d} \}$ into any arbitrary orthonormal basis $\{ \ket{1'},\ldots,\ket{d'} \}$, hence our object is the unitary group of dimension $d$. We know that for dimension $d$ the unitary group $\mathcal{U}(d)$ is a $d^2$-parameter group. This means that in order to cover all unitary transformations we must find a transformation that contains at least $d^2$ parameters (if a parameterization involves more than $d^2$ parameters it also contains redundancies which is undesirable).\\
Regarding the basis $\{ \ket{1},\ldots,\ket{d} \}$, we first find that $d$ parameters can be embedded in global phases for each vector, i.e. $\{ e^{i \alpha_1} \ket{1},\ldots,e^{i \alpha_d} \ket{d} \}$. This corresponds to the product $ \prod_{l=1}^{d} \mbox{exp}(i P_l \lambda_{l,l})$ of our parameterization (\ref{Uc}). The notion behind the left part of (\ref{Uc}) is to pairwise embed two parameters in the distinctive transformations
\begin{align}
\label{distrans}
\Lambda_{m,n}=\mbox{exp} \left( i P_n \lambda_{n,m} \right) \mbox{exp} \left( i \sigma_{m,n} \lambda_{m,n} \right)
\end{align}
containing two parameters $\lambda_{n,m}$ and $\lambda_{m,n}$. These transformations perform the following operations: The term $\mbox{exp} \left( i \sigma_{m,n} \lambda_{m,n} \right)$ generates a rotation in the two-dimensional subspace spanned by the vectors $\ket{m}$ and $\ket{n}$, while $\mbox{exp} \left( i P_n \lambda_{n,m} \right)$ adds a relative phase between the vector components of the rotated vectors. Note that $\mathcal{U}(2)$ is actually a $4$-parameter group, however, as will be proven later, the neglected two parameters would only lead to redundancies. In order to parameterize $\mathcal{U}(d)$ all terms $\Lambda_{m,n}$ have to be taken into account. There are $\begin{pmatrix} d\\2 \end{pmatrix}=d(d-1)/2$ ways how two vectors of $\{ \ket{1},\ldots,\ket{d} \}$ can be combined, which corresponds to the $d(d-1)/2$ generalized anti-symmetric $\sigma$-matrices. One opportunity to involve all terms $\Lambda_{m,n}$ is given by building the product $\prod_{m=1}^{d-1} \left(\prod_{n=m+1}^{d} \Lambda_{m,n}  \right)$, which is clearly not unique. We now have embedded further $2 \times d(d-1)/2 = d^2-d$ parameters, which in sum with the $d$ global phases gives $d^2$ parameters $\lambda_{m,n}$ in total. For convenience, these parameter $\lambda_{m,n}$ are gathered in a $d \times d$ 'parameterization' matrix
\begin{align}
\left(
  \begin{array}{ccc}
    \lambda_{1,1} & \cdots & \lambda_{1,d} \\
    \vdots & \ddots & \vdots \\
    \lambda_{d,1} & \cdots & \lambda_{d,d} \\
  \end{array}
\right) \ .
\end{align}
In this convention the diagonal entries represent global phase transformations, the upper right are related to rotations, while the lower left are relative phases (with respect to the basis $\{\ket{1},\ldots,\ket{d}\}$).\\
We have thus constructed a $d^2$ parameter set of unitary operators. It remains to prove that this construction covers the entire unitary group $\mathcal{U}(d)$, which is equivalent to showing that for any $U$ there exists a $U_C$ such that $U_C^{\dagger} U=\mathbbm{1}$.\\
\textbf{Proof:}\\
Let $U=\sum_{r,s=1}^{d} a_{r,s} \ket{r}\bra{s}$ be an arbitrary unitary operator, i.e. $\sum_{i=1}^{d} a_{m,i}^{*}a_{n,i}=\sum_{i=1}^{d} a_{i,m}^{*}a_{i,n}=\delta_{mn}$. The conjugate transpose of $U_C$ is
\begin{align}
U_C^{\dagger}=\left[ \prod_{l=1}^{d} \mbox{exp}(-i P_{d+1-l} \lambda_{d+1-l,d+1-l})\right]\ \cdot \left[\prod_{m=1}^{d-1} \left(\prod_{n=1}^{m}  \Lambda_{d-m,d+1-n}^{\dagger}  \right) \right] \ ,
\end{align}
implying that $\Lambda_{1,2}^{\dagger}$ acts first on $U$. For $U'=\Lambda_{1,2}^{\dagger} U=\sum_{r,s=1}^{d} a'_{r,s} \ket{r}\bra{s}$ this leads to
\begin{align}
\label{proofa}
a'_{1,s}&=\cos(\lambda_{1,2})a_{1,s}-e^{-i\lambda_{2,1}}\sin(\lambda_{1,2})a_{2,s} \ , \\
a'_{2,s}&=\sin(\lambda_{1,2})a_{1,s}+e^{-i\lambda_{2,1}}\cos(\lambda_{1,2})a_{2,s} \ .
\end{align}
For $d>2$, all other components are unchanged, i.e. $a'_{r,s}=a_{r,s}$ for $r>2$. As can easily be confirmed, $a'_{2,1}$ can always be made zero via certain parameters $\lambda_{1,2}$ and $\lambda_{2,1}$. If $a_{1,1}$ and $a_{2,1}$ both are zero, both parameters $\lambda_{1,2}$ and $\lambda_{2,1}$ can be chosen freely; if only $a_{1,1}=0$ we choose $\lambda_{1,2}=\frac{\pi}{2}$ and for $a_{2,1}=0$ we take $\lambda_{1,2}=0$. If none of both is zero then $a'_{2,1}$ vanishes for $\lambda_{2,1}$ and $\lambda_{1,2}$ obeying
\begin{align}
arg(e^{-i\lambda_{2,1}}a_{2,1})&=arg(-a_{1,1}) \ ,\\
\tan(\lambda_{1,2})&=\frac{|a_{2,1}|}{|a_{1,1}|} \ ,
\label{proofb}
\end{align}
which is achievable in any case with $\lambda_{2,1}\in [0,2\pi]$ and a $\lambda_{1,2}\in [0,\frac{\pi}{2}]$. In the same way the component $a''_{3,1}$ of $U''=\Lambda^{\dagger}_{1,3}U'$ can be made zero for $d>2$. By proceeding in this way for $d>2$, we can attain $a_{r,s}=0$ for all components with $r>s$ via $\prod_{m=1}^{d-1} \left(\prod_{n=1}^{m}  \Lambda_{d-m,d+1-n}^{\dagger}  \right) $. Then, if $a_{r,1}=0$ for all $r>1$ it follows $|a_{1,1}|=1$ due to unitarity $(\sum_{i=1}^{d} a_{i,1}^{*}a_{i,1}=1)$ of $U_P=\left[\prod_{m=1}^{d-1} \left(\prod_{n=1}^{m}  \Lambda_{d-m,d+1-n}^{\dagger}  \right) \right]U$. Moreover, since $U_P$ also satisfies $\sum_{i=1}^{d} a_{1,i}^{*}a_{1,i}=1$ we can infer that $a_{1,r}=0$ for all $r>1$. When taking into account the unitarity constraints $\sum_{i=1}^{d} a_{m,i}^{*}a_{n,i}=\sum_{i=1}^{d} a_{i,m}^{*}a_{i,n}=\delta_{mn}$ for all rows and columns we conclude that $U_P$ has the diagonal form
\begin{align}
U_P=\sum_{r=1}^{d} a_{r,r} \ket{r}\bra{r}
\end{align}
with $a_{r,r}$ obeying $|a_{r,r}|=1$, which is a complex number of magnitude $1$, i.e. $a_{r,r}=e^{i\alpha_r}$. The choice $\lambda_{r,r}=\alpha_r$ then yields  $U_C^{\dagger}U=\left[ \prod_{l=1}^{d} \mbox{exp}(-i P_{d+1-l} \lambda_{d+1-l,d+1-l})\right] U_P =\mathbbm{1}$, which was to be proven.
\section{Parameterization of density matrices with rank $k$}
Now we use our parameterization of $\mathcal{U}(d)$ to formulate density matrices. Any density matrix $\rho$ acting on $\mathcal{H}=\mathbb{C}^d$ can be written in the form
\begin{align}
\rho=\sum_{n=1}^{k} p_n \ket{\Psi_n} \bra{\Psi_n}
\end{align}
with $p_n\geq0$ and $\sum_{n=1}^{k}p_n=1$ where $k\leq d$ is the rank of $\rho$ and $\{ \ket{ \Psi_1 },\ldots,\ket{ \Psi_k } \}$ are orthonormal vectors. For $k=1$, i.e. pure states we only have one $p_1=1$. Without loss of generality, the coefficients $\{ p_n \}$ of an arbitrary state of rank $k>1$ or smaller can be expressed by means of $k-1$ real parameters $\theta_i \in [0,\frac{\pi}{2}]$ as
\begin{align}
p_1&=cos^2\theta_1\\
p_n&=cos^2\theta_n \prod_{i=1}^{n-1} sin^2 \theta_i \hspace{1cm} \forall \, n \ : \ 1<n<k\\
p_k&=\prod_{i=1}^{k-1} sin^2\theta_i \ .
\end{align}
Any desired set of $k$ orthonormal vectors $\{\ket{\Psi_1},\ldots,\ket{\Psi_k} \}$ can be constructed out of $\{ \ket{1},\ldots,\ket{k} \}$ by applying $U_C$. Thus any $\rho$ with rank $k$ can be parameterized via
\begin{align}
\label{pdensitymatrix}
\rho=\sum_{n=1}^{k} p_n U_C \ket{n} \bra{n} U_C^{\dagger} \ .
\end{align}
A parameterization of density matrices via diagonal elements and unitary transformations can in principle be achieved with any parameterization of the unitary group $\mathcal{U}(d)$ (see for instance \cite{ByrdSlater,TilmaApp,Bruening}). However, the composite form of $U_C$ makes it easy to identify and eliminate all redundant parameters. The first observation is that the diagonal entries $\lambda_{n,n}$ are redundant, since they cancel out in the outer product, i.e.
\begin{align}
\left[ \prod_{l=1}^{d} \mbox{exp}(i P_l \lambda_{l,l})\right] \ket{n} \bra{n} \left[ \prod_{l=1}^{d} \mbox{exp}(-i P_{d+1-l} \lambda_{d+1-l,d+1-l})\right] = \ket{n} \bra{n} \hspace{0.5cm} \forall n \in\{1,..,d\} \ .
\end{align}
The second observation is that for density matrices of rank $k<d$ we can eliminate further parameters when the composite parameterization is used. As can easily be seen $\rho$ is independent of all parameters $\lambda_{m,n}$ where both $m>k$ \underline{and} $n>k$. Thus, it suffices to utilize
\begin{align}
\rho=\sum_{n=1}^{k} p_n U_{CD} \ket{n} \bra{n} U_{CD}^{\dagger}
\end{align}
with
\begin{align}
\label{UCD}
U_{CD}=\prod_{m=1}^{k} \left(\prod_{n=m+1}^{d} \mbox{exp} \left( i P_n \lambda_{n,m} \right) \mbox{exp} \left( i \sigma_{m,n} \lambda_{m,n} \right)  \right) \ ,
\end{align}
where the index $m$ is only running from $1$ to $k$ instead of $1$ to $d-1$. Using the introduced representation in terms of the parameterization matrix, this is $U_C$ with
\begin{align}
& \left(
  \begin{array}{ccccccc}
    0 & \lambda_{1,2} & \cdots & \lambda_{1,k+1} & \cdots & \lambda_{1,d} \\
    \lambda_{2,1} & \ddots & \ddots & \vdots & \ddots & \vdots \\
    \vdots & \ddots& 0 &  \lambda_{k,k+1} & \cdots & \lambda_{k,d} \\
         \lambda_{k+1,1} & \cdots & \lambda_{k+1,k}   & 0 & \cdots & 0 \\
     \vdots & \ddots & \vdots   & \vdots & \ddots & \vdots\\
     \lambda_{d,1} & \cdots & \lambda_{d,k}   & 0 & \cdots & 0 \\
   \end{array}
\right)
\begin{array}{c}
  \left. {\begin{array}{c}
  \\
  \\
  \\
    \end{array}} \right\} k \hspace{7mm} \\
    \\
    \left. {\begin{array}{c}
  \\
  \\
  \\
  \end{array}} \right\} d-k
\end{array} \\
& \hspace{0.6cm} \underbrace{ \hspace{3.1cm} }_{k} \hspace{7mm} \underbrace{ \hspace{2.8cm} }_{d-k} \hspace{2cm} . \nonumber
\end{align}
Consequently, any density matrix  of rank $k$ or smaller acting on $\mathbb{C}^d$ can be expressed with a maximum of $2dk-k^2-1$ parameters ($k-1$ parameters $\theta_{i}$ and $k(2d-k-1)$ parameters $\lambda_{m,n}$). The extremal scenarios are pure states with $2(d-1)$ parameters and full rank density matrices with $d^2-1$ parameters.

\section{Parameterization of $k$-dimensional subspaces}
In the previous section we have shown that any set of $k$ orthonormal vectors can be constructed (up to global phases) with $k(2d-k-1)$ parameters. Now, we prove that even less, namely $2k(d-k)$ parameters are necessary to construct an orthonormal basis of an arbitrary $k$-dimensional subspace of $\mathbb{C}^d$. Consider a general subspace spanned by $k$ orthonormal vectors $\{\ket{\Psi_1},\ldots,\ket{\Psi_k}\}$ in a $d$-dimensional complex Hilbert space $\mathcal{H}=\mathbb{C}^d$. Let $a_{m,n}$ be the coefficients of $\ket{\Psi_n}$ in the complete basis $\{\ket{1},\ldots,\ket{d}\}$ of $\mathcal{H}=\mathbb{C}^d$, i.e. $\ket{\Psi_n}=\sum_{m=1}^{d}a_{m,n} \ket{m}$. A different basis of the same subspace then is of course given by the orthonormal set of vectors
\begin{align}
 \mbox{\Large \{} \ket{\Psi_1},\ldots,\ket{\Psi_{k-2}},&\cos(\lambda_{k-1,k})\ket{\Psi_{k-1}}-e^{(i\lambda_{k,k-1})} \sin(\lambda_{k-1,k}) \ket{\Psi_k},\\
&\sin(\lambda_{k-1,k})\ket{\Psi_{k-1}}+e^{(i\lambda_{k,k-1})} \cos(\lambda_{k-1,k}) \ket{\Psi_k} \mbox{\Large \}} \ ,
\end{align}
which is $\{\Lambda'_{k-1,k} \ket{\Psi_1},\ldots,\Lambda'_{k-1,k} \ket{\Psi_k} \}$ where $\Lambda'_{k-1,k}$ is given as in (\ref{distrans}) but with $P_l=\ket{\Psi_l}\bra{\Psi_l}$ and $\sigma_{m,n}=-i\ket{\Psi_m} \bra{\Psi_n} + i \ket{\Psi_n} \bra{\Psi_m}$. The $(k-1)th$ vector of this set is
\begin{align}
& \cos(\lambda_{k-1,k})\ket{\Psi_{k-1}}-e^{(i\lambda_{k,k-1})} \sin(\lambda_{k-1,k}) \ket{\Psi_k}\\
=& \sum_{m=1}^{d}\left( \cos(\lambda_{k-1,k})a_{m,k-1}-e^{(i\lambda_{k,k-1})} \sin(\lambda_{k-1,k}) a_{m,k} \right) \ket{m} \ ,
\end{align}
whose k$th$ coefficient $\cos(\lambda_{k-1,k})a_{k,k-1}-e^{(i\lambda_{k,k-1})} \sin(\lambda_{k-1,k}) a_{k,k}$ can be made zero analogously to (\ref{proofa})-(\ref{proofb}).
In this way, with $\{ \prod_{n=1}^{k-1} \Lambda'_{n,k} \ket{\Psi_1},\ldots,\prod_{n=1}^{k-1} \Lambda'_{n,k} \ket{\Psi_k} \}$ a basis of the subspace can be obtained where all k$th$ coefficients of all vectors except the last one are zero. This can then be repeated for all rows $<k$, such that with $\{ \prod_{m=2}^{k} \prod_{n=1}^{m-1} \Lambda_{n,m}' \ket{\Psi_1},\ldots,\prod_{m=2}^{k} \prod_{n=1}^{m-1} \Lambda_{n,m}' \ket{\Psi_k} \}$ we arrive at an orthonormal basis, let us say $\{ \ket{\Psi'_1},\ldots,\ket{\Psi'_k} \}$, of the same subspace where all coefficients $a_{m,n}=0$ for all $m$ with $n<m\leq k$. Such a set of vectors however, can be constructed (up to global phases) from $\{\ket{1},\ldots,\ket{k}\}$ via $\{U_{CS}\ket{1},\ldots,U_{CS}\ket{k}\}$ where
\begin{align}
\label{Ucs}
U_{CS}=\prod_{m=1}^{k} \left(\prod_{n=k+1}^{d} \mbox{exp} \left( i P_n \lambda_{n,m} \right) \mbox{exp} \left( i \sigma_{m,n} \lambda_{m,n} \right)  \right) \ .
\end{align}
Here, only $2k(d-k)$ parameters $\lambda_{m,n}$ are involved since the index $n$ of the second product $\prod$ lies within $k < n \leq d$. In terms of the parameterization matrix, this is $U_C$ with
\begin{align}
& \left(
  \begin{array}{ccccccc}
    0 & \cdots & 0 & \lambda_{1,k+1} & \cdots & \lambda_{1,d} \\
    \vdots & \ddots & \vdots & \vdots & \ddots & \vdots \\
    0& \cdots& 0 &  \lambda_{k,k+1} & \cdots & \lambda_{k,d} \\
         \lambda_{k+1,1} & \cdots & \lambda_{k+1,k}   & 0 & \cdots & 0 \\
     \vdots & \ddots & \vdots   & \vdots & \ddots & \vdots\\
     \lambda_{d,1} & \cdots & \lambda_{d,k}   & 0 & \cdots & 0 \\
   \end{array}
\right)
\begin{array}{c}
  \left. {\begin{array}{c}
  \\
  \\
  \\
    \end{array}} \right\} k \hspace{7.2mm}\\
    \\
    \left. {\begin{array}{c}
  \\
  \\
  \\
  \end{array}} \right\} d-k
\end{array} \\
& \hspace{0.6cm} \underbrace{ \hspace{3.1cm} }_{k} \hspace{7mm} \underbrace{ \hspace{2.8cm} }_{d-k} \hspace{2cm} . \nonumber
\end{align}
\textbf{Proof:}\\
 The proof is completely analogous to the one in section 2 and will therefore not be given in detail. One only has to take an arbitrary set of vectors $\{\ket{\Psi'_1},\ldots,\ket{\Psi'_k}\}$ with the mentioned properties, for which via $\{ U^{\dagger}_{CS} \ket{\Psi'_1},\ldots,U^{\dagger}_{CS} \ket{\Psi'_k} \}=\{ \sum_{m=1}^{d} a_{m,1}\ket{m},\ldots,\sum_{m=1}^{d} a_{m,k}\ket{m} \}$ it is possible to attain $|a_{m,n}|=\delta_{m,n}$ for all $m \in \left[ 1, \ldots , d \right]$ and $n \in \left[ 1, \ldots , k \right]$ - QED.
\section{Application: Optimization of lower bounds of entanglement measures}
The $m$-concurrence introduced in Ref.~\cite{HH2} constitutes a building block of entanglement measures for multipartite systems of arbitrary dimension (see also Ref.~\cite{HHK1}) with simple computable lower bounds. These bounds are not invariant under local unitary transformations. Optimization procedures can be realized efficiently via the previously introduced parameterization of the unitary group.
\subsection{Optimal lower bounds}
\begin{figure*}
\centering (a)\includegraphics[scale=0.31]{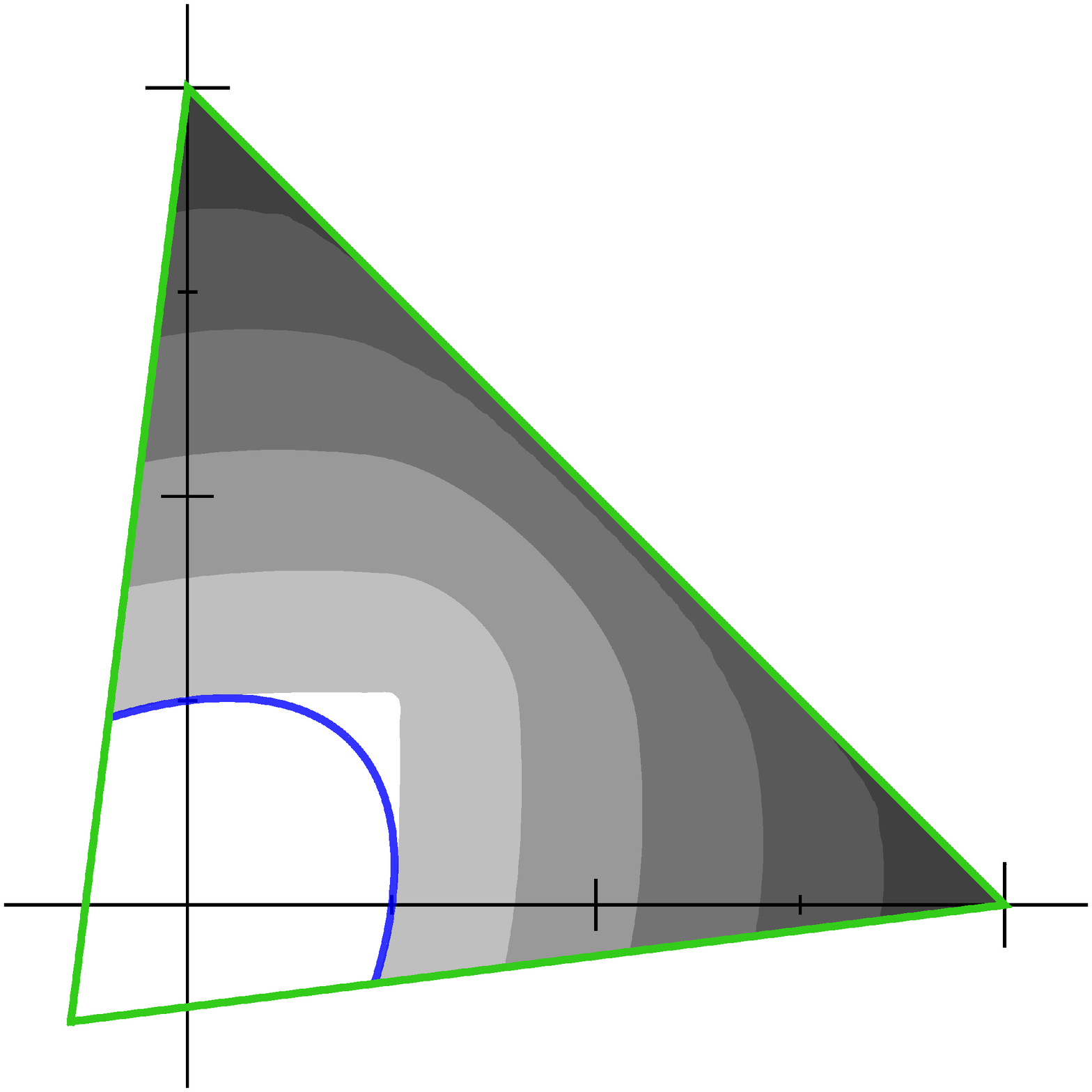}
(b)\includegraphics[scale=0.31]{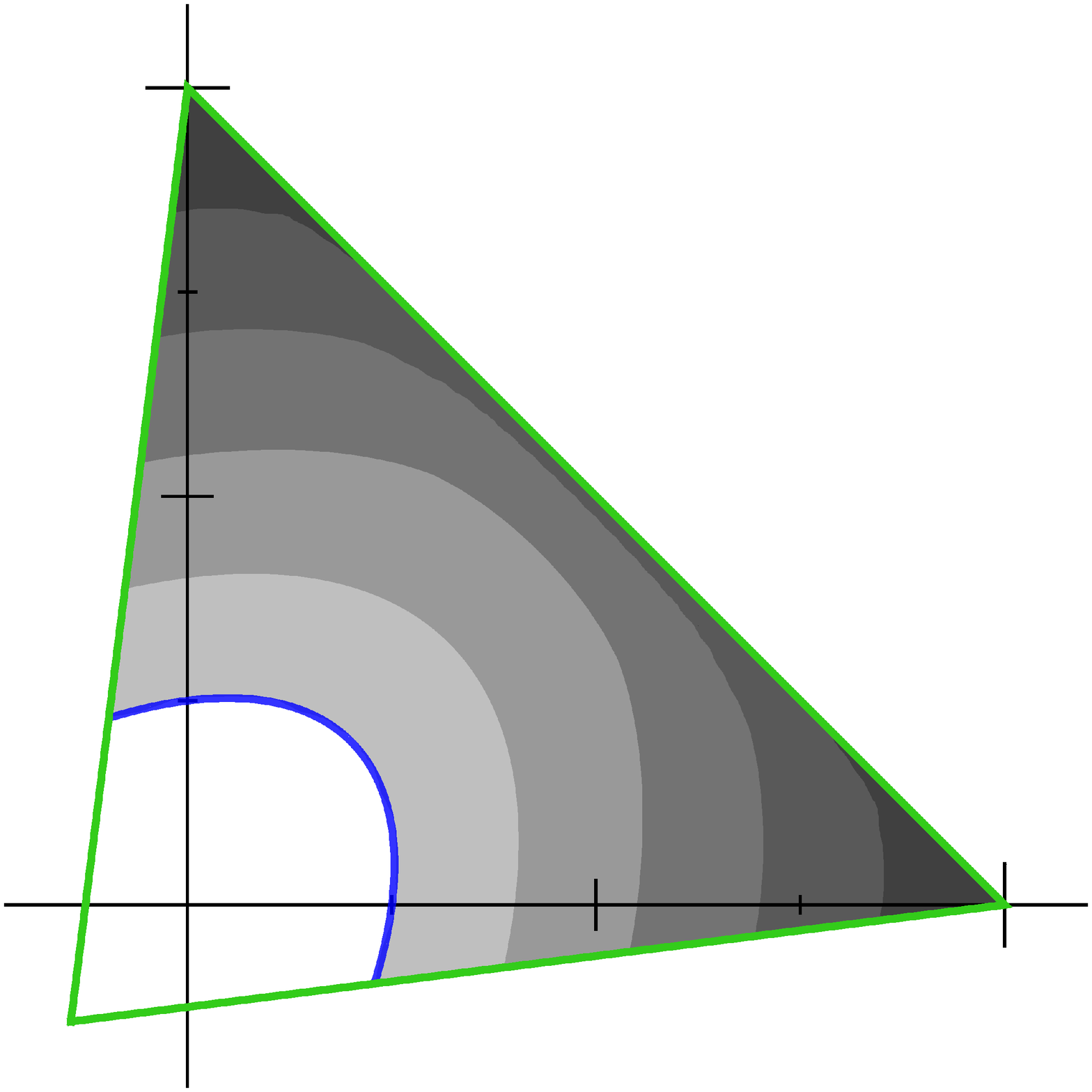}
\setlength{\unitlength}{1cm}%
  \begin{picture}(0, 0)(0,0)
  \put(-8.3,0.6){\fontsize{12}{14}\selectfont \makebox(0,0)[]{$1$ \strut}}
\put(-0.6,0.6){\fontsize{12}{14}\selectfont \makebox(0,0)[]{$1$ \strut}}
\put(-14,6.3){\fontsize{12}{14}\selectfont \makebox(0,0)[]{$1$ \strut}}
\put(-6.3,6.3){\fontsize{12}{14}\selectfont \makebox(0,0)[]{$1$ \strut}}

\put(-8.0,1.6){\fontsize{12}{14}\selectfont \makebox(0,0)[]{$\alpha$ \strut}}
\put(-0.3,1.6){\fontsize{12}{14}\selectfont \makebox(0,0)[]{$\alpha$ \strut}}
\put(-13.2,6.7){\fontsize{12}{14}\selectfont \makebox(0,0)[]{$\beta$ \strut}}
\put(-5.5,6.7){\fontsize{12}{14}\selectfont \makebox(0,0)[]{$\beta$ \strut}}
\end{picture}
  \caption{Contour plots of the lower bounds $B(\rho)$ and $B_{opt}(\rho)$ of the normalized $m$-concurrence $\frac{C_m(\rho)}{C_m(\ket{\Psi_{1}}\bra{\Psi_{1}})}$ for the set of states $\rho=\alpha \ket{\Psi_{1}}\bra{\Psi_{1}}+\beta \ket{\Psi_{2}}\bra{\Psi_{2}}+\frac{1-\alpha-\beta}{9}\mathbbm{1}$ constructed with two mutually orthogonal maximally entangled states $\ket{\Psi_{1}}=\frac{1}{\sqrt{3}}\left(\ket{11}+\ket{22}+\ket{33}\right)$ and $\ket{\Psi_{2}}=\frac{1}{\sqrt{3}}\left(\ket{12}+\ket{23}+\ket{31}\right)$ of $\mathbb{C}^3 \otimes \mathbb{C}^3$ (bipartite qutrit system) combined with uncolored noise $\mathbbm{1}$. All values of the parameters $\alpha$ and $\beta$ within the green triangle correspond to density matrices (positive semidefinite). Density matrices within the blue ellipse are positive under partial transposition (PPT criterion). In the grey shaded areas the bounds of the $m$-concurrence are nonzero. The grey scale corresponds to the value of the bounds lying in the range of $0$ and $1$. The shades of grey are related to an increment of $0.2$ starting from $0$. The left picture (a) illustrates the contour plot of the bounds computed according to equation (\ref{mcon2}) without optimization. The right picture (b) illustrates the contour plot of the numerically optimized bounds (\ref{optibound}) using our parameterization for $\mathcal{U}(3)$ and the Nelder-Mead method. Without optimization (a), the lower bounds  are zero for some states with negative partial transposition. After the numerical optimization (b), the lower bounds of all states that are negative under partial transposition are greater than zero. As can also be seen, the shapes of the contours of (b) differ considerably from (a), and indicate an improvement of the lower bounds.}
  \label{figure1}
\end{figure*}

First we introduce the $m$-concurrence for bipartite quantum systems in $\mathbb{C}^d\otimes\mathbb{C}^d$. The linear entropy $S_L(\rho):=\frac{d}{d-1}(1-\text{Tr}(\rho^2))$ of the reduced density matrices $\rho_{A/B}:=\text{Tr}_{B/A}(\rho)$ of a bipartite pure state $\ket{\psi} \in \mathbb{C}^d\otimes\mathbb{C}^d$ can be expressed as
\begin{align}
\frac{2(d-1)}{d}S_L(\rho_A)&=\sum_{k_A<l_A}\sum_{k_B<l_B} \text{Tr}(\ket{\psi}\bra{\psi}\sigma_{k_A,l_A}\otimes\sigma_{k_B,l_B}(\ket{\psi}\bra{\psi})^*\sigma_{k_A,l_A}\otimes\sigma_{k_B,l_B})\\ &=:C_m^2(|\psi\rangle\langle\psi|) \ .
\end{align}
Via a convex roof extension one can generalize the m-concurrence $C_m^2(|\psi\rangle\langle\psi|)$ to mixed states
\begin{align}
\label{mcon1}
C_m^2(\rho):&=\inf_{\{p_i,|\psi_i\rangle\}}\sum_i p_i C^2_m(|\psi_i\rangle\langle\psi_i|)\\
&= \inf_{\{p_i,|\psi_i\rangle\}}\sum_i  p_i \sum_{k_A<l_A}\sum_{k_B<l_B}  \text{Tr}(\ket{\psi_i}\bra{\psi_i}\sigma_{k_A,l_A}\otimes\sigma_{k_B,l_B}(\ket{\psi_i}\bra{\psi_i})^*\sigma_{k_A,l_A}\otimes\sigma_{k_B,l_B}) \,. \nonumber
\end{align}
As the infimum of this expression cannot straightforwardly be computed, it is of great importance to find strong lower bounds. Those can be obtained with
\begin{align}
C_m^2(\rho)\geq\sum_{k_A<l_A}\sum_{k_B<l_B}\inf_{\{p_i,|\psi_i\rangle\}}\sum_i p_i \text{Tr}(\ket{\psi_i}\bra{\psi_i}\sigma_{k_A,l_A}\otimes\sigma_{k_B,l_B}(\ket{\psi_i}\bra{\psi_i})^*\sigma_{k_A,l_A}\otimes\sigma_{k_B,l_B}) \nonumber
\end{align}
by exploiting that the individual infima are known (see Ref.~\cite{mintert05,wootters98})
\begin{align}
\inf_{\{p_i,|\psi_i\rangle\}}\sum_i p_i \text{Tr}(\ket{\psi_i}\bra{\psi_i}\sigma_{k_A,l_A}\otimes\sigma_{k_B,l_B}(\ket{\psi_i}\bra{\psi_i})^*\sigma_{k_A,l_A}\otimes\sigma_{k_B,l_B}) =X^2_{k_A,l_A,k_B,l_B}\, ,
\end{align}
with
\begin{align}
\label{bounds}
X_{k_A,l_A,k_B,l_B}:=\max[2\max[\{x^i_{k_A,l_A,k_B,l_B}\}]-\sum_ix^i_{k_A,l_A,k_B,l_B},0]
\end{align}
where $\{x^i_{k_A,l_A,k_B,l_B}\}$ are the square roots of the eigenvalues of
\begin{align}
\rho \ \sigma_{k_A,l_A}\otimes\sigma_{k_B,l_B} \ \rho^* \ \sigma_{k_A,l_A}\otimes\sigma_{k_B,l_B} \ .
\end{align}
In summary we have obtained the following bound
\begin{align}
\label{mcon2}
C^2_m(\rho)\geq \sum_{k_A<l_A}\sum_{k_B<l_B}X^2_{k_A,l_A,k_B,l_B} =: B^2(\rho) \ .
\end{align}
This bound is not invariant under local unitaries and thus can be optimized with an appropriate choice $U_A\otimes U_B$. Consequently, the optimal lower bound for the $m$-concurrence for bipartite systems is given by
\begin{align}
B^2_{opt}(\rho):=\max \left[\sum_{k_A<l_A}\sum_{k_B<l_B}X^2_{k_A,l_A,k_B,l_B} \right] \, ,
\label{optibound}
\end{align}
where the eigenvalues of $\rho \ \sigma_{k_A,l_A}\otimes\sigma_{k_B,l_B} \ \rho^* \ \sigma_{k_A,l_A}\otimes\sigma_{k_B,l_B}$ are replaced by the eigenvalues of
\begin{align}
\label{EWU}
U_A \otimes U_B \ \rho \ U^{\dagger}_A\otimes U^{\dagger}_B \sigma_{k_A,l_A} \otimes \sigma_{k_B,l_B} U_A^* \otimes U_B^* \rho^* U^T_A \otimes U^T_B \sigma_{k_A,l_A} \otimes \sigma_{k_B,l_B}
\end{align}
and the maximum is taken over all $U_A,U_B \in \mathcal{U}(d)$. In many cases, this problem can be solved numerically by parameterzing $U_A$, $U_B$ and utilizing numerical methods such as 'Nelder-Mead' \cite{NelderMead}, 'simulated annealing' \cite{SA} or 'differential evolution' \cite{DE}. In order to remove redundant parameters we exploit that
\begin{align}
&\mbox{EV}\left(U_A \otimes U_B \rho U^{\dagger}_A\otimes U^{\dagger}_B \sigma_{k_A,l_A} \otimes \sigma_{k_B,l_B} U_A^* \otimes U_B^* \rho^* U^T_A \otimes U^T_B \sigma_{k_A,l_A} \otimes \sigma_{k_B,l_B}\right)\\
=&\mbox{EV}\left(\rho \left(U^{\dagger}_A  \sigma_{k_A,l_A} U_A^*\right)  \otimes \left(U^{\dagger}_B  \sigma_{k_B,l_B}  U_B^* \right) \rho^* \left(U^T_A \sigma_{k_A,l_A} U_A  \right) \otimes \left(U^T_B \sigma_{k_B,l_B} U_B \right) \right)
\label{optend}
\end{align}
where EV stands for the set of eigenvalues. If we now insert $U_A^{\dagger}$ and $U_B^{\dagger}$ in parameterized form (\ref{Uc}), i.e. $U_A^{\dagger}=U_{C,A}$ and $U_B^{\dagger}=U_{C,B}$, and let them act on the $\sigma$-matrices, it is easy to see that all diagonal entries $\lambda_{m,m}$ of both unitaries cancel out. Consequently, we can reduce (\ref{optibound}) to a $2(d^2-d)$ dimensional global optimization problem. (If we would insert $U_A=U_{C,A}$ and $U_B=U_{C,B}$, only one diagonal entry $\lambda_{m,m}$ of each $U_C$ could generally be set zero.) An illustrative example, where the bounds of the $m$-concurrence are optimized in this way using the Nelder-Mead method \cite{NelderMead} for convex combinations of two mutually orthogonal maximally entangled states and uncolored noise is given in Fig.~\ref{figure1}. With optimization the bounds are greater than zero for all states detected by the partial transposition criterion (PPT).

For multipartite systems we can use the same principle (for a detailed introduction see Refs.~\cite{HH2,HHK1}). First we introduce the set of all bipartitions $\mathcal{B}=\{ \left( \alpha|\beta \right) \}$ of a given $n$-partite system. Here, $\alpha$ denotes a union of subsystems in the first part of the bipartition and $\beta$ is its complement. The dimensions of the corresponding complex Hilbert spaces $\mathcal{H}_{\alpha}$ and $\mathcal{H}_{\beta}$ are $d_\alpha$ and $d_\beta$, respectively. In this way, the general definition of the $m$-concurrence also valid for multipartite systems is
\begin{align*}
C^2_{m}(\rho)=\inf_{\{p_i,|\psi_i\rangle\}}\sum_{i}p_i \sum_{\mathcal{B}}\sum_{k_A<l_A}\sum_{k_B<l_B}\text{Tr}(\ket{\psi_i}\bra{\psi_i}\sigma_{k_A,l_A}^{\alpha\beta}\otimes\sigma_{k_B,l_B}^{ \alpha\beta }(\ket{\psi_i}\bra{\psi_i})^*\sigma_{k_A,l_A}^{ \alpha\beta}\otimes\sigma_{k_B,l_B}^{\alpha\beta})
\end{align*}
where $\{ \sigma_{k_A,l_A}^{\alpha \beta} \}$ and $\{ \sigma_{k_B,l_B}^{\alpha \beta} \}$ are the generalized anti-symmetric $\sigma$-matrices defined with respect to the bipartition $\left( \alpha|\beta \right)$, i.e. acting on $\mathcal{H}_{\alpha}$ and $\mathcal{H}_{\beta}$, respectively. Consequently, according to (\ref{mcon1}) - (\ref{mcon2})
\begin{align}
C^2_{m}(\rho) \geq \sum_{\mathcal{B}}\sum_{k_A<l_A}\sum_{k_B<l_B} \left(X_{k_A,l_A,k_B,l_B}^{\alpha \beta}\right)^2 \ ,
\end{align}
where $X_{k_A,l_A,k_B,l_B}^{\alpha \beta}$ is defined as in (\ref{bounds}) but with the eigenvalues of
\begin{align}
\rho \ \sigma_{k_A,l_A}^{\alpha\beta}\otimes\sigma_{k_B,l_B}^{ \alpha\beta } \ \rho^* \ \sigma_{k_A,l_A}^{ \alpha\beta}\otimes\sigma_{k_B,l_B}^{\alpha\beta} \ .
\end{align}
By proceeding analogously to (\ref{optibound}) - (\ref{optend}), this lower bound can be optimized by replacing $\rho$ by $U_A^{\alpha \beta} \otimes U_B^{\alpha \beta} \rho U_A^{\dagger \alpha \beta} \otimes U_B^{\dagger \alpha \beta}$ for each bipartition. In this sense, the optimal lower bound is given by
\begin{align}
B^2_{opt}(\rho):=\max \left[\sum_{\mathcal{B}}\sum_{k_A<l_A}\sum_{k_B<l_B}\left(X^{\alpha \beta}_{k_A,l_A,k_B,l_B}\right)^2 \right] \, ,
\end{align}
over all $U_A^{\alpha \beta} \in \mathcal{U}(d_{\alpha})$ and $U_B^{\alpha \beta} \in \mathcal{U}(d_{\beta})$.
\subsection{Distillation}
If entanglement is used as a resource it is often required that the system is in a pure maximally entangled state. States that can be transformed into such maximally entangled states via LOCC (local operations and classical communication) are called distillable. It was proven by Horodecki \textit{et al.} in \cite{Horodeckibound} that all distillable states have an entangled $\mathbb{C}^2\otimes\mathbb{C}^2$ subsystem (see also Ref.~\cite{Bruss}). As the bounds for the $m$-concurrence are exact in those systems it suffices to optimize one of the terms in the sum of (\ref{optibound}) to investigate distillability since all terms are local-unitarily related. Thus, a state $\rho$ is distillable only if
\begin{align}
\max X_{1,2,1,2}^2>0 \ ,
\end{align}
where the maximum is taken over all $U_A \in \mathcal{U}(d)$ and $U_B \in \mathcal{U}(d)$. According to (\ref{optend}), the value of $X_{1,2,1,2}$ is a function of the eigenvalues of
\begin{align}
\rho \left(U^{\dagger}_A  \sigma_{1,2} U_A^*\right)  \otimes \left(U^{\dagger}_B  \sigma_{1,2}  U_B^* \right) \rho^* \left(U^T_A \sigma_{1,2} U_A  \right) \otimes \left(U^T_B \sigma_{1,2} U_B \right) \ .
\end{align}
These eigenvalues, however, are completely determined by the $2\times 2$ dimensional subspace spanned by the tensorproducts of the vectors $U^{\dagger}_A \ket{1}$ and $U^{\dagger}_A \ket{2}$ with $U^{\dagger}_B \ket{1}$ and $U^{\dagger}_B \ket{2}$. Consequently, for $U^{\dagger}_A$ and $U^{\dagger}_B$ we can take $U^{\dagger}_A = U_{CS,A}$ and $U^{\dagger}_B = U_{CS,B}$ defined as in (\ref{Ucs}) with $k=2$, where each transformation only depends on $4d-8$ parameters $\lambda_{m,n}$. In this way, the question of whether a state $\rho$ is distillable or not, can efficiently be clarified via a (numerical) optimization algorithm with a reduced number of parameters. (Note that in contrast to a naive parameterization of $\mathcal{U}(d)$, the number of parameters is linear in $d$ instead of quadratic. As distillability is generally studied for high-dimensional $n$-copy states, i.e. $\rho^{\otimes n}$, this reduction is of great importance for numerical tractability.)

Multipartite systems can be treated equivalently for each bipartition. Hence, for a fixed bipartition $4(d_{\alpha}+d_{\beta})-16$ variables have to be optimized. However, in this case, if the unitary operators related to a fixed bipartition are not locally implementable with respect to the subsystems, also unlockable bound entanglement can appear (see also Refs.~\cite{HHHKS1,HH3}). The set of locally distillable states can be determined by restricting all transformations to unitaries which are local with respect to all subsystems.
\section{Summary}
In this paper we introduced a parameterization of the unitary group $\mathcal{U}(d)$
which beside its simple form has the advantage that redundancies can easily be identified and removed in several cases. The efficiency for the representation of density matrices of rank $k$ and $k$-dimensional subspaces in $\mathbb{C}^d$ was shown. We found that only $2dk-k^2-1$ real parameters are necessary to parameterize density matrices of rank $k$ since we were able to discard all irrelevant parameters related to transformations in uninvolved subspaces and invariant phases beforehand. For the construction of an
orthonormal basis of any $k$-dimensional subspace of $\mathbb{C}^d$
even less parameters are needed, namely $2k (d-k)$, due to the unitary equivalence of basis vectors within the subspace.\\
Furthermore, examples of the usefulness of the parameterization with respect to a multipartite entanglement measure (referred to as the $m$-concurrence) and distillability were given. We described how lower bounds can be derived for this entanglement measure. The bounds were obtained by the observation
that the linear entropy of reduced density matrices can be rewritten by an
operator sum, where each operator acts in a two-dimensional
subspace. We showed how these bounds can be optimized and how invariant parameters can be removed when our parameterization is utilized. Finally, we revealed that further parameters can be discarded if only the distillability of quantum states is of interest. We found that only a number of variables linear in the dimensions of the subsystems has to be optimized to solve this problem.\\
We believe that the parameterization presented in this paper is advantageous with respect to the tractability of a variety of high-dimensional optimization procedures in quantum information as well as for other problems in quantum theory.\\
\\
\textbf{Acknowledgements:}
We thank Andreas Gabriel, Marcel Meyer and Robert W. Johnson for helpful remarks and Petre Dita for informing us about his related works Refs.~\cite{Dita1,Dita2}. Christoph Spengler and Marcus Huber acknowledge financial support from the Austrian FWF, Project P21947N16. Beatrix C. Hiesmayr acknowledges the fellowship MOEL 428.\\
\\

\end{document}